

\input phyzzx

\catcode`@=11

\def\figitem#1{\r@fitem{Fig.~#1:}}
\tolerance=1000

%
\newtoks\KUNS   \KUNS={0000}
\newtoks\HETH   \HETH={00/00}
\newtoks\monthyear
\monthyear={\monthname,\ \number\year}
\Pubnum={KUNS~\the\KUNS\cr HE(TH)~\the\HETH\cr hep-ph/9207260}
\def\p@bblock{\begingroup \tabskip=\hsize minus \hsize
   \baselineskip=1.5\ht\strutbox \topspace-2\baselineskip
   \halign to\hsize{\strut ##\hfil\tabskip=0pt\crcr
   \the\Pubnum\cr \the\monthyear\cr }\endgroup}
\def\bftitlestyle#1{\par\begingroup \titleparagraphs
     \iftwelv@\fourteenpoint\else\twelvepoint\fi
   \noindent {\bf #1}\par\endgroup }
\def\title#1{\vskip\frontpageskip \bftitlestyle{#1} \vskip\headskip }
%
%

%
\def\journal#1&#2(#3){\begingroup \let\journal=\dummyj@urnal
    \unskip, \sl #1\unskip~\bf\ignorespaces #2\rm
    (\afterassignment\j@ur \count255=#3) \endgroup\ignorespaces }
\def\andjournal#1&#2(#3){\begingroup \let\journal=\dummyj@urnal
    \sl #1\unskip~\bf\ignorespaces #2\rm
    (\afterassignment\j@ur \count255=#3) \endgroup\ignorespaces }
\def\andvol&#1(#2){\begingroup \let\journal=\dummyj@urnal
    \bf\ignorespaces #1\rm
    (\afterassignment\j@ur \count255=#2) \endgroup\ignorespaces }
\def\NP{Nucl.~Phys. }
\def\PR{Phys.~Rev. }
\def\PRL{Phys.~Rev.~Lett. }
\def\PL{Phys.~Lett. }
\def\PTP{Prog.~Theor.~Phys. }
\def\CMP{Commun.~Math.~Phys.}

%

\def\abs#1{{\left\vert #1 \right\vert}}
\def\bigcdot{\hbox{
             \vbox{\vskip-0.1cm\hbox{\seventeenrm .}\vskip0.1cm}}}
\def\cdot{\mathchar"2201}
\def\Half{{1 \over 2}}
\def\QB{Q_{\rm B}}
\def\T{{\rm T}}
\def\e{e}
\def\calL{{\cal L}}
\def\calO{{\cal O}}
\def\calD{{\cal D}}
\def\Linv{\calL_{\rm inv}}
\def\Lgauge{\calL_{\rm gauge}}
\def\dB{\delta_{B}}
\def\cb{\overline{c}}
\def\sumpm{\sum_\pm}
\def\d{\partial}

\def\deriv#1{\left(\partial/\partial #1\right)}
\def\Rpart{\overrightarrow{\partial}}
\def\Lpart{\overleftarrow{\partial}}
\font\circlefont=lcircle1 
\def\blob{\hbox{\raise 3pt \vbox{
          \hbox{\thinspace\circlefont \char"76}}}}
\def\circle{\hbox{\raise 3pt \vbox{
            \hbox{\thinspace\thinspace\circlefont \char"66}}}}
\def\hyp{\hbox{-}}

\catcode`@=12


\REF\Wilson{K.G.~Wilson \journal \PR &D10 (74) 2445.} 

\REF\KOsuppl{T.~Kugo and I.~Ojima \journal \PTP~Suppl. &66 (79).} 

\REF\KugoCONF{T.~Kugo \journal \PL &83B (79) 93.} 

\REF\NGPHOTON{
R.~Ferrari and L.E.~Picasso \journal \NP &B31 (71) 316; \nextline
R.A.~Brandt and Ng~Wing-Chiu \journal \PR &D10 (74) 4198.}

\REF\HataRES{H.~Hata \journal \PTP &67 (82) 1607;
\andvol &69 (83) 1524.} 

\REF\tHooftCONF{G.~'t\thinspace Hooft \journal \NP &B138 (78) 1;
\andjournal \NP &B153 (79) 141.} 

\REF\HataPG{H.~Hata \journal \PL &143B (84) 171.} 

\REF\HataKugoPG{H.~Hata and T.~Kugo \journal \PR &D32 (85) 938.}

\REF\WittenTFT{E.~Witten \journal \CMP &117 (88) 353.} 

\REF\DelJar{R.~Delbourgo and P.D.~Jarvis
\journal J.~Phys. &A15 (82) 611.} 

\REF\ParisiSourlas{
G.~Parisi and N.~Sourlas \journal \PRL &43 (79) 744.} 

\REF\PolyWieg{
A.~Polyakov and P.B.~Wiegmann \journal \PL &131B (83) 121;\nextline
P.B.~Wiegmann \journal \PL &141B (84) 217; \andvol &142B (84) 173.}

\REF\SuzuShima{
T.~Suzuki and K.~Shimada \journal \PTP &69 (83) 1537;\nextline
T.~Suzuki \journal \PTP &69 (83) 1827.} 

\REF\tHooftABEL{G.~'t\thinspace Hooft
\journal \NP &B190[FS3] (81) 455.} 

\REF\ABELMONOPOLE{A.S.~Kronfeld, G.~Schierholz and U.-J. Wiese
\journal \NP &B293 (87) 461;\nextline
T.~Suzuki \journal \PTP &81 (89) 752;\nextline
J.~Smit and A.J.~van~der~Sijs \journal \NP &B355 (91) 603.}

\REF\KugoUehara{T.~Kugo and S.~Uehara \journal \NP &B197 (82) 378.}

\REF\HataIkkiII{H.~Hata and I.~Niigata, in preparation.} 

\REF\SFT{E.~Witten \journal \NP &B268 (86) 2530;\nextline
H.~Hata, K.~Itoh, T.~Kugo, H.~Kunitomo and K.~Ogawa
\journal \PR &D34 (86) 2360.} 

\REF\Planck{
I.~Klebanov and L.~Susskind \journal \NP &B309 (88) 175;\nextline
J.~Atick and E.~Witten \journal \NP &B310 (88) 291
} 


\KUNS={1146}
\HETH={92/09}

\titlepage

\title{Color Confinement, Abelian Gauge and Renormalization
Group Flow}

\author{Hiroyuki HATA\foot{\rm Bitnet address: hata@jpnyitp}
and Ikki NIIGATA\foot{\rm Bitnet address: ikki@jpnyitp}
}

\address{Department of Physics,~Kyoto University \break
                            Kyoto~606,~JAPAN}

\abstract{
Under the assumption that the color charge can be
written in a BRST exact form,
the color confinement mechanism proposed by Kugo and Ojima (KO)
explains
the confinement of any colored particles including dynamical quarks and
gluons.
This mechanism, however, is known to break down in the Abelian gauge
which treats the maximal Abelian subgroup of the gauge group in a
special manner.
In order to study whether the failure of the KO mechanism is particular
only to the Abelian gauge or whether this failure occurs in a wide
class of gauges including the ordinary Lorentz type gauge,
we carry out a renormalization group study of the $SU(2)$ gauge theory
in the gauge fixing space.
Our gauge fixing space consists of four distinct regions that are not
connected with each other by renormalization group flows,
and we find that the Abelian gauge is {\it infrared unstable}
in three regions which include the Lorentz type gauge.
This suggests that the failure of the KO mechanism is a phenomenon
which occurs only in the Abelian gauge.
We also find that the Lorentz gauge is infrared stable.
}

\endpage
\chapter{Introduction}

The subject of this paper is color confinement, a long-standing
problem in Yang-Mills theory. In most of the literature this problem
has been discussed in terms of the Wilson loop\rlap.\refmark{\Wilson}
Namely, if the Wilson loop expectation value obeys the area decay law,
it implies that the potential between a {\it static} quark and
anti-quark pair is a linear one and therefore the pair can never be
infinitely separated.
Although the Wilson loop has such a simple and intuitive interpretation
and its area decay has been observed in lattice gauge theories,
it is useless when there are {\it dynamical} quark fields
in the system and cannot explain the confinement of
colored particles other than quarks (\eg, gluons).
In this paper we shall consider a different approach proposed by Kugo
and Ojima\rlap,\refmark{\KOsuppl,\KugoCONF}
which treats the confinement of dynamical quarks as well as gluons.

Let us recapitulate the confinement mechanism of Kugo and Ojima
(referred to as the KO mechanism hereafter).
We consider Yang-Mills theory with Lorentz type covariant gauge
fixing (see ref.~[\KOsuppl] for details). The Lagrangian is
$$
\calL = \Linv
+ i\dB\left\{ \cb\bigcdot \left(\d_\mu A_\mu
- {\alpha \over 2}B \right) \right\}\ ,
\eqn\eqL
$$
where $\Linv=-(1/4g^2)F_{\mu \nu }^2 + \calL_{\rm matter}$ is the
gauge invariant Lagrangian and the BRST transformation $\dB$ is
defined as usual by
$$
\eqalign{
&\dB A_\mu = D_\mu c \equiv  \d_\mu c + A_\mu \times c\ ,
\qquad \dB c = - \Half c\times c\ ,\cr
&\dB \cb = iB\ ,\qquad \dB B = 0 \ ,\cr
}\eqn\eqBRSTR
$$
for the gauge fields $A_\mu$, the Faddeev-Popov (FP) ghost and the
anti-ghost fields, $c$ and $\cb$, the multiplier field $B$,
and similarly for matter fields.
The key quantity in the KO mechanism is the conserved color current
$N_\mu$ defined by
$$
N_\mu = -i\dB \left(D_\mu \cb\right)=\left\{ \QB, D_\mu \cb\right\}\ ,
\eqn\eqN
$$
where $\QB$ is the BRST charge.
$N_\mu$ actually generates a global color rotation since if we make
on $\calL$ \eqL\ a local gauge transformation with parameter
$\epsilon (x)$, $\delta A_\mu =D_\mu \epsilon$ and
$\delta (c,\cb,B)=(c,\cb,B)\times \epsilon$,
we obtain
$\delta \calL= N_\mu \!\bigcdot\d_\mu \epsilon
+ \hbox{[total div. term]}$ (note that $[\delta ,\dB]=0$).
Therefore $N_\mu$ has its origin only from the gauge part $\dB(*)$ of
$\calL$ and is written as a BRST exact form.
The ordinary Noether color current $J_\mu$ containing the matter
fields is related to the present
$N_\mu$ by $J_\mu =N_\mu  + (1/g^2)\d_\nu F_{\nu \mu }$
using the equation of motion of $A_\mu$.

Kugo-Ojima's observation is as follows. If the color current $N_\mu$
contains no massless one particle component and the integration
$\int d^3x N_{\mu =0}$ has a well-defined meaning, then the color
charge $Q^a$ can be expressed as a BRST exact form:
$$
Q^a = \left\{ \QB, \int d^3x \left(D_0\cb\right)^a \right\}\ .
\eqn\eqQA
$$
This BRST exact form of the color charge implies color confinement
since we can easily show using eq.~\eqQA\ that any asymptotic state
belonging to a non-trivial representation of color $Q^a$
necessarily belongs to a quartet representation of the BRST algebra
and hence is {\it unphysical} and unobservable\rlap.\refmark{\KOsuppl}
(In Appendix A, we reproduce the proof of the above statement given in
ref.~[\KOsuppl].)

The condition that there be no massless one particle component
in the current $N_\mu$ is {\it not} satisfied in perturbation theory:
the two-point function $\VEV{\T N_\mu  A_\nu }$ has the general
expression
$$
i\int d^4x \e^{ip\cdot x}\VEV{\T N_\mu^a(x) A_\nu^b(0)} =
\left(\delta^{ab} + u^{ab}(p^2)\right){p_\mu p_\nu \over p^2}
- u^{ab}(p^2)g_{\mu \nu } \ ,
\eqn\eqNA
$$
where $u^{ab}(p^2)=O(\hbar)$ comes from loop diagrams and we have
$u\equiv 0$ when the gauge group is $U(1)$.
The massless pole part $p_\mu p_\nu /p^2$ is the contribution of the
$B$ field ($B \propto \d_\mu A_\mu$) present in
$N_\mu =\d_\mu B + \hbox{(composites)}$.
However, this massless pole disappears and the color confinement
is realized if the condition due to Kugo and Ojima,
$$
\delta^{ab} + u^{ab}\left(p^2 \to 0\right) = 0\ ,
\eqn\eqKOCOND
$$
is satisfied non-perturbatively (of course we are assuming that
$N_\mu$ does not develop a new massless pole which is absent in
perturbation theory).

Although the KO confinement mechanism has not been proved in QCD,
we shall mention two investigations which support
this possibility. One is the interpretation of the KO mechanism as the
restoration of a local gauge symmetry. It is well known that in
QED with Lorentz type gauge fixing the photon field $A_\mu$ can be
regarded as the Nambu-Goldstone boson corresponding to the
spontaneous breakdown of the local gauge symmetry
$\delta A_\mu (x)= \d_\mu \lambda (x)= a_\mu$ with
$\lambda (x)=a_\mu x_\mu$\rlap.\refmark{\NGPHOTON}
In ref.~[\HataRES], it was shown that the KO mechanism
can be interpreted as the {\it restoration} of a similar kind of
local gauge symmetry which is spontaneously broken in perturbation
theory.
This implies that the KO mechanism is associated
with the {\it disordered} vacuum which contains large fluctuations of
the gauge field.
Since the area decay law of the Wilson loop is also considered
to be a consequence of the disordered gauge field
vacuum\rlap,\refmark{\tHooftCONF}
we see that the KO mechanism and the Wilson loop area decay are
phenomena akin to each other although their direct relationship is not
yet known.

The other topic concerning the KO mechanism is the ``pure-gauge model"
discussed in refs.~[\HataPG, \HataKugoPG].
{}From the observation that the KO mechanism has an intimate relationship
with large gauge field fluctuation in the direction of local gauge
transformation, the authors of refs.~[\HataPG, \HataKugoPG] devised
a toy model which is obtained from the Yang-Mills theory by
restricting the gauge field to the pure-gauge configuration
$A_\mu (x)=g^{-1}(x)\d_\mu g(x)$.
Since the gauge field is pure-gauge, the action consists solely of the
gauge part (\ie, the gauge-fixing and the FP ghost terms)
written as a BRST exact form $\dB(*)$ (it is a kind of topological
field theory\refmark{\WittenTFT} in modern terminology).
Although this pure-gauge model has no physical degrees of freedom,
its color current $N_\mu$ \eqN\ contains a massless pole in
perturbation theory, and it is a non-trivial dynamical problem whether
this massless pole disappears non-perturbatively.

This pure-gauge model is a kind of non-linear $\sigma$-model since the
elementary field is not a vector field $A_\mu$
but is the gauge group element $g(x)$, although it differs from the
ordinary non-linear $\sigma$-model in that the present model contains
{\it fourth} derivatives in the action and that it has the FP-ghosts
and the BRST symmetry.
However, owing to these differences the pure-gauge model in
four dimensions resembles in several aspects the non-linear
$\sigma $-model in two dimensions.
In particular, if we adopt a special gauge called
the $OSp(4/2)$-symmetric gauge\refmark{\DelJar}
which admits a superspace formulation,
the pure-gauge model in four dimensions is in the
Parisi-Sourlas\refmark{\ParisiSourlas} sense exactly equivalent to
the chiral model in two dimensions.
Concretely the following relation holds between the two-point functions
of the two theories:
$$
\int d^4x \e^{ip\cdot x}\VEV{\T N_\mu^a(x) A_\nu^b(0)}_{\rm 4\hyp D} =
-2\pi \int d^2x\e^{ip\cdot x}
\VEV{\T A_\mu^a(x)A_\nu^b(0)}_{\rm 2\hyp D}\ ,
\eqn\eqEQUIV
$$
where the left hand side (right hand side) is the quantity in the
$OSp(4/2)$-symmetric pure-gauge model in four dimensions (chiral model
in two dimensions), and the 4-momentum $p_\mu$ and the indices
$\mu$ and $\nu$ on the left hand side have components only in the
two dimensional part where the chiral model of the right hand side
lives.
Since it is known that dynamical mass generation occurs in the
chiral model in two dimensions\refmark{\PolyWieg} and the two-point
function of $A_\mu =g^\dagger\d_\mu g$ on the right hand side of
eq.~\eqEQUIV\ has no massless singularity at $p^2=0$,
this equivalence implies that the KO confinement condition
eq.~\eqKOCOND\ is actually satisfied in the pure-gauge model with
$OSp(4/2)$-symmetric gauge fixing.

So far so good. However, there is a serious objection to the KO
mechanism raised by Suzuki and Shimada\rlap.\refmark{\SuzuShima}
They took a special gauge called the Abelian
gauge\rlap.\refmark{\tHooftABEL}\foot{
The Abelian gauge has been used for discussing confinement
(by the Wilson loop) on the basis of the monopole configuration.
See, for example, refs.~[\tHooftABEL] and [\ABELMONOPOLE].
}
When the gauge group is $SU(2)$, this gauge is given by the following
gauge fixing functions $F^\pm$ and $F^3$ (which should be replaced
by $\d_\mu A_\mu$ in the Lorentz gauge):
$$
\eqalign{
&F^\pm[A] = \left(\d_\mu  \pm iA_\mu^3\right)A_\mu^\pm\ , \cr
&F^3[A] = \d_\mu A_\mu^3\ ,
}\eqn\eqABELGAUGE
$$
where the $\pm$ component is defined by
$$
\calO^\pm=\left(\calO^1 \pm i\calO^2\right)/\sqrt{2}\ .
\eqn\eqPMDEF
$$
The meaning of eq.~\eqABELGAUGE\ is as follows.
First, $F^\pm$ partially fixes the local $SU(2)$ symmetry but
completely preserves the local $U(1)_3$ symmetry, the local rotation
symmetry around the 3rd axis in $SU(2)$
(note that $F^\pm$ transforms covariantly under the local $U(1)_3$).
At this stage we have a $U(1)$ gauge theory having $A_\mu^3$ as the
``photon" field. The Lorentz type gauge fixing $F^3$ then fixes
the remaining local $U(1)_3$ symmetry.
For a general gauge group the Abelian gauge is similarly
defined by replacing the derivative in the Lorentz gauge
with the covariant derivative with respect to the maximal Abelian
subgroup.

In this Abelian gauge let us consider the confinement of the $Q^{a=3}$
charge in the KO mechanism. We see immediately that the KO
mechanism cannot work here. This is because as far as the $U(1)_3$
is concerned the non-Abelian gauge theory with Abelian gauge fixing is
nothing but a $U(1)$ gauge theory with Lorentz type gauge fixing.
Concretely, in the Abelian gauge the BRST exact color current of
$U(1)_3$ is given by $N_\mu^3=\d_\mu B^3$ and we have the exact
two-point function $\hbox{F.T.}\VEV{TB^3 A_\mu^3}=p_\mu /p^2$.
Therefore in eq.~\eqNA\ we have $u^{33}(p^2)\equiv 0$.
Does this suggest that the KO mechanism cannot work even in the Lorentz
gauge? Or is the Abelian gauge exceptional?

The purpose of this paper is to study using the renormalization group
(RG) whether the KO confinement mechanism fails in other gauges than
the Abelian gauge.
We take $SU(2)$ as the gauge group and consider a class of gauges
defined by the gauge fixing function
$F^\pm[A] = \left(\d_\mu  \pm i\xi A_\mu^3\right)A_\mu^\pm$
with a parameter $\xi$ which includes both the Lorentz gauge
($\xi =0$) and the Abelian gauge ($\xi =1$),
and study the one-loop RG flow of $\xi$ towards the infrared (IR)
direction.
The Abelian gauge is one of the fixed points
of the RG since it is protected by the symmetry under the local
$U(1)_3$ rotation with angle $\epsilon^3(x)= a_\mu x_\mu$.
If the Abelian gauge $\xi =1$ is IR unstable, it implies that the IR
properties of the Abelian gauge $\xi \equiv 1$ may be completely
different from other $\xi \not=1$ gauges and the failure of the KO
mechanism may therefore be a property particular only to the
Abelian gauge.

In order to carry out the RG analysis we have to consider a wider
class of gauges defined by five parameters including the above $\xi$.
The calculation has been done both for the Yang-Mills theory (without
the matter fields) and the pure-gauge model.
We find that the Abelian gauge is IR unstable in most of the parameter
space. We also find that the Lorentz gauge is IR stable.

One might think it strange that we are going to consider the RG flow in
the gauge fixing space as a means of studying color confinement
which should hold independently of the choice of gauge.
We expect that confinement by the KO mechanism holds in all gauges
which preserve (a part of) the global color symmetry.
Note, however, that we cannot restrict our discussion to the gauge
invariant (physical) sector of the Yang-Mills theory when considering
the KO mechanism.
This is evident from the way confinement is realized in the KO
mechanism (color non-singlets become {\it unphysical}),
and from the fact that the KO confinement condition of eq.~\eqKOCOND\
is not (at least naively) a gauge invariant quantity.
What we are going to study in this paper is in a sense the gauge fixing
independence of the KO confinement mechanism.

The organization of the rest of this paper is as follows. In the
next section a general class of gauges for studying the RG is defined.
In Sec.~3, we first present the $\beta$-functions for the parameters
and then study the RG flow for both the pure-gauge model and the
Yang-Mills theory.
The final section is devoted to a summary and discussion.
In the Appendix we present the proof that the BRST exact expression
of the color charge implies color confinement.

\chapter{The model}

In the BRST quantization of gauge theories, the most general
form of the gauge fixing term is given by\refmark{\KugoUehara}
$$
\Lgauge = -i\dB G \ ,
\eqn\eqSGAUGE
$$
where $G$ carries the ghost number $-1$ and is a hermitian function
of $A_\mu$, $c$ ,$\cb$, $B$ and possibly the matter fields.
In this paper we are interested in a class of gauges
which is characterized by a number of parameters and contains as
special cases the Abelian gauge and the Lorentz gauge.
If we impose only the global $U(1)_3$ rotation symmetry around the 3rd
axis, the most general gauge fixing function $G$ consisting of
operators of dimension three or less is
$$
\eqalign{
G &= \sumpm\cb^\mp \left(
\left(\d_\mu \pm i\xi_\pm A^3_\mu \right)A^\pm_\mu
- {\alpha \over 2}B^\pm \right) \cr
&+ \cb^3\left( \d_\mu A^3_\mu  + a\left(A^3_\mu \right)^2
 + b\abs{A^+_\mu }^2 - {\beta \over 2}B^3 + M^2\right) 
+ \sumpm (\pm)\eta_\pm \cb^3\cb^\pm c^\mp + \zeta c^3\cb^+\cb^-\ ,\cr
}\eqn\eqGENERALG
$$
where $\alpha$, $\beta$, $a$, $b$, $M^2$ and $\zeta$ are
real parameters,
and the complex parameters $\xi_\pm$ and $\eta_\pm$ satisfy
$\xi_\mp=\xi_\pm^*$ and $\eta_\mp=\eta_\pm^*$.
However, eq.~\eqGENERALG\ contains too many parameters to analyze.
We therefore make the restriction
$a=b=M^2=0$, and $\xi_\pm =\xi$ (real) and $\eta_\pm=\eta$ (real)
by imposing the symmetry under ``charge conjugation",
$$
\calO^\pm \rightarrow \calO^\mp,\qquad \calO^3 \rightarrow - \calO^3\ ,
\eqn\eqPARITY
$$
for any fields $\calO=A_\mu , c, \cb, B$\rlap.\foot{
The transformation \eqPARITY\ is expressed in terms of the Lie algebra
valued fields $\calO\equiv \sum_a \left(\sigma_a/2i\right)\calO^a$ as
$\calO \rightarrow  \sigma_1\calO \sigma_1$.}
Note that the invariance of $G$ under \eqPARITY\ implies the
invariance of $\dB G$ since the BRST transformation is commutative
with \eqPARITY.

Now our gauge fixing function is given by
$$
G = \sumpm \cb^\mp \left(F^\pm[A] - {\alpha \over 2}B^\pm \right)
+ \cb^3\left(F^3[A] - {\beta \over 2}B^3\right)
+\alpha \left(
\eta \sumpm(\pm)\cb^3\cb^\pm c^\mp +\zeta c^3\cb^+\cb^- \right) \ ,
\eqn\eqG
$$
with 
$$
\eqalign{
&F^\pm[A]= \left(\d_\mu \pm i\xi A^3_\mu \right)A^\pm_\mu  \ , \cr
&F^3[A] = \d_\mu A^3_\mu  \ . \cr
}
\eqn\eqF
$$
Compared with eq.~\eqGENERALG\
we have redefined $\eta$ and $\zeta$ in eq.~\eqG\ by
multiplying them by $\alpha$ for later convenience.
The BRST transformation formula in the $(\pm ,3)$ basis is
$$
\eqalign{
&\dB A^\pm_\mu =
\left(\d_\mu \pm iA^3_\mu \right)c^\pm \mp iA^\pm_\mu c^3\ ,\cr
&\dB A^3_\mu = \d_\mu c^3 +
i\left(A^+_\mu c^- - A^-_\mu c^+\right)\ ,\cr
&\dB c^\pm = \mp i c^3c^\pm \ ,\cr
&\dB c^3 = -i c^+ c^- \ ,\cr
&\dB \cb^a = i B^a \ ,\qquad \dB B^a = 0 .
}
\eqn\eqDBPMTHREE
$$
For the gauge fixing function $G$ of eq.~\eqG, the gauge fixing
lagrangian \eqSGAUGE, after eliminating the $B$ fields, is given
explicitly by
$$
\eqalign{
&\Lgauge = {1\over \alpha }\abs{F^+[A]}^2
+ {1\over 2\beta }\left(F^3[A]\right)^2
-i\sumpm \d_\mu \cb^\mp \d_\mu  c^\pm -i\d_\mu \cb^3\d_\mu  c^3 \cr
&+\sumpm \Biggl[
\cb^\mp\left\{
i\xi \left( \abs{A^+_\mu }^2 - \left(A^3_\mu \right)^2 \right)
-(\pm)\left(({\alpha \over \beta }\eta +1)F^3[A]
+ (1+\xi )A^3_\mu \Rpart_{\!\!\mu } \right)\right\} c^\pm \cr
&-i\xi \cb^\mp\left(A^\pm_\mu \right)^2 c^\mp
+(\pm)\cb^\mp\left\{
(1+\zeta )F^\pm[A]+(1-\xi )A^\pm_\mu \Rpart_{\!\!\mu }\right\}c^3 \cr
&- (\pm)\cb^3\left(\eta F^\pm[A]
- \Lpart_{\!\!\mu } A^\pm_\mu \right)c^\mp \Biggr]\cr
&- \alpha (1+\zeta )\eta \sumpm\cb^3 c^3\cb^\pm c^\mp
- \alpha \left(
\zeta +{\alpha \over \beta }\eta^2\right)\cb^+ \cb^- c^+ c^- \ .\cr
}\eqn\eqOURLGAUGE
$$

In our gauge with parameters $(\xi ,\eta ,\zeta ,\alpha ,\beta )$
the Abelian gauge corresponds to the subspace $\xi =1$,
where the system has an invariance under the ``local" $U(1)_3$
rotation with angle $\epsilon^3(x)=a_\mu x_\mu$,
$$
A^3_\mu \rightarrow A^3_\mu
+ \d_\mu \epsilon^3\ ,\quad \calO^3\rightarrow \calO^3\ ,\quad
\calO^\pm\rightarrow \e^{\mp i\epsilon^3}\calO^\pm \ ,
\eqn\eqUTHREEROT
$$
where $\calO^3=(c,\cb,B)^3$ and $\calO^\pm=(A_\mu ,c,\cb,B)^\pm$.
The KO mechanism breaks down for any $\eta$ and $\zeta$
when $\xi =1$.

On the other hand the Lorentz gauge having the full global $SU(2)$
symmetry corresponds to the two points,
$(\xi ,\eta ,\zeta ,\alpha /\beta )=(0,0,0,1)$ and $(0,-1,-1,1)$.
These two points are related to each other by the FP-ghost conjugation
$c\rightarrow \cb$, $\cb\rightarrow -c$.
Other restrictions on $\Lgauge$ \eqOURLGAUGE\ protected by symmetries
are
$$
\eqalign{
\hbox{i)}&\ \eta =0 \quad \hbox{by the symmetry}\quad
\cb^3 \rightarrow  \cb^3 + \overline{\theta } \ ,\cr
\hbox{ii)}&\ \zeta =-1 \quad \hbox{by the symmetry}\quad
c^3 \rightarrow  c^3 + \theta  \ ,\cr
\hbox{iii)}&\ \xi =0, \eta =\zeta , \alpha /\beta =1 \quad
\hbox{by the global $SU(2)$ symmetry} \ ,\cr
\hbox{iv)}&\ (\xi ,\eta ,\zeta ,\alpha /\beta )=(0,-1/2,-1/2,1) \quad
\hbox{by iii) and the FP-ghost conjugation symmetry} \ ,\cr
}\eqn\eqRESTRICTION
$$
where $\overline{\theta }$ and $\theta$ are constant
Grassmann numbers. The FP-ghost conjugation\foot{
$\Lgauge$ \eqOURLGAUGE\ is closed under the FP-ghost conjugation
only when the condition
$\alpha \left(\eta (1-\xi )-\zeta -1\right)+\beta (1-\xi )^2=0$
is satisfied. It is defined by
$c^\pm\rightarrow \cb^\pm$, $\cb^\pm\rightarrow -c^\pm$,
$c^3\rightarrow \cb^3/(1-\xi )$, $\cb^3\rightarrow -(1-\xi )c^3$,
and in the resultant $\Lgauge$ the parameters $\zeta$ and $\eta$ are
replaced by
$-\eta (1-\xi )-1$ and $-(1+\zeta )/(1-\xi )$, respectively,
while other parameters are kept invariant.}
symmetry characterizing the point
$(\xi ,\eta ,\zeta ,\alpha /\beta )=(0,-1/2,-1/2,1)$
may be replaced by the $OSp(4/2)$
symmetry\refmark{\DelJar, \HataKugoPG} in the case of
the pure-gauge model
(note that $OSp(4/2)$ is not a symmetry of $\Linv$).

In the following we shall work in the Euclidean space,
and our pure Yang-Mills theory is described by the path-integral
$$
Z_{\rm Yang\hyp Mills} =
\int \calD A_\mu  \calD c \calD\cb \exp\left(-S\right) \ ,
\eqn\eqZ
$$
with the total action $S$,
$$
S = {1\over g^2}\int d^4x
\left( {1\over 4}\left(F_{\mu \nu }\right)^2 + \Lgauge \right) \ ,
\eqn\eqS
$$
where $g$ is the gauge coupling constant.

Finally the pure gauge model corresponding to the gauge fixing \eqG\
is obtained from the Yang-Mills system \eqZ\ as follows:
First redefine $\cb$ and $B$ as
$(\cb, B)_{\rm pure-gauge} = (1/g^2)(\cb, B)_{\rm Yang-Mills}$
and then take the limits $g\rightarrow 0$ and $\alpha ,
\beta \rightarrow \infty$ for the gauge coupling
constant $g$ and the gauge parameters $\alpha$ and $\beta$
in such a way that $t^2\equiv g^2\alpha$ and
$u^2\equiv g^2\beta$ are kept finite.
By taking the limit $g\rightarrow 0$ the gauge field $A_\mu$
becomes constrained to the pure-gauge configuration
$A_\mu (x)=g^\dagger(x)\d_\mu g(x)$
($g(x)$: $SU(2)$ matrix) and the system is now described by the
path-integral
$$
Z_{\rm pure\hyp gauge} = \int \calD g \calD c \calD\cb \exp\left\{
- \int d^4x\Lgauge\left(A_\mu =g^\dagger\d_\mu g\right)
\Big\vert_{(\alpha ,\beta )\ {\rm replaced\ by}\ (t^2,u^2)}
\right\} \ .
\eqn\eqZPG
$$
%
%

\chapter{Renormalization group analysis}

\section{$\beta$-functions}

The one-loop renormalization calculation of our pure Yang-Mills system
with non-linear gauge fixing \eqOURLGAUGE\ is straightforward but
rather lengthy since we have only the $U(1)_3$ symmetry of the $SU(2)$
gauge group and the gauge fixing term \eqOURLGAUGE\ contains
complicated interaction vertices absent in the linear gauge.
In fact, we have to calculate more than fifty diagrams to determine the
renormalization constants.

Here we present the $\beta$-function $\beta_X$ for the coupling
constants $X=g, \xi , \zeta$ etc. in the Yang-Mills system \eqZ.
The $\beta$-function here is defined by multiplying the ordinary
definition by
$8\pi^2$: $\beta_X=8\pi^2\deriv{\ln\mu }X\vert_{\rm bare\ fixed}$
with $\mu$ being the renormalization point. We have
$$
\eqalign{
&\beta_g = - {11\over 3}g^3 \ ,\cr
&\beta_\xi /g^2= {1\over 4}(1-\xi )\left\{
3\xi^2 + (5\alpha +9)\xi +\alpha \zeta
- \alpha \eta (1-\xi )\right\} \ ,\cr
&\beta_\eta /g^2= {1\over 4}\eta \left\{
3\xi (\xi +3) + {12\over \alpha }\xi^2
+ \alpha \left[ 5\xi +6\zeta +1+2\eta (\xi +4\zeta +3)\right]
+ 3\beta (1-\xi )^2 \right\}\ ,\cr
&\beta_\zeta /g^2= \alpha (1+\zeta )\left\{
\zeta + 3\left({\xi \over \alpha }\right)^2
+ 2\eta (\xi +\zeta )\right\} \ ,\cr
&\beta_r/g^2= -{3\over 2}r\xi (\xi +3) - {\alpha \over 2}(1-\xi )^2
- {3\over \beta }\xi^2
- \alpha r\left\{ \Half(5\xi +2\zeta -1)+\eta (3\xi +2\zeta -1)
- 2r\eta^2 \right\}, \cr
&\beta_\alpha /g^2= \alpha \left\{ {13\over 3}-{3\over 2}\xi (\xi +1)
- {\alpha \over 2}\left[3\xi +2\zeta +1 +
2\eta (\xi +2\zeta +1)\right] -{\beta \over 2}(1-\xi )^2
\right\} - 3\xi^2 \ ,\cr
&\beta_\beta /g^2= \beta \left\{{13\over 3} + 3\xi
- \alpha (1-\xi )(1+2\eta ) \right\}- 2\alpha^2\eta^2 \ ,\cr
}\eqn\eqBETAFUN
$$
where $r$ is defined by $r\equiv \alpha /\beta$.

The $\beta$-functions in the pure-gauge model of our non-linear
gauge are obtained by taking the limits $g\rightarrow 0$ and
$\alpha,\beta \rightarrow \infty$ with $t^2=g^2\alpha$
and $u^2=g^2\beta$ kept fixed.
They are much simpler than eq.~\eqBETAFUN:
$$
\eqalign{
&\beta_\xi /t^2 = {1\over 4}(1-\xi )\left\{
5\xi  + \zeta - \eta (1-\xi )\right\} \ ,\cr
&\beta_\eta /t^2 = {1\over 4}\eta \left\{
5\xi +6\zeta +1 +{3\over r}(1-\xi )^2 + 2\eta (\xi +4\zeta +3)
\right\} \ ,\cr
&\beta_\zeta /t^2 = (1+\zeta )\left\{
\zeta + 2\eta (\xi +\zeta ) \right\} \ ,\cr
&\beta_r/t^2 = - r\left\{
\Half(5\xi +2\zeta -1) + \eta (3\xi +2\zeta -1) \right\}
            - \Half(1-\xi )^2 + 2r^2\eta^2 \ ,\cr
&\beta_{(1/t^2)} = {1\over 2r}(1-\xi )^2 +{1\over 2}(1+3\xi )+\zeta
+ \eta (1+\xi +2\zeta ) \ ,\cr
&\beta_{(1/u^2)} = r(1-\xi )(1+2\eta )+ 2r^2\eta^2 \ .\cr
}\eqn\eqPGBETAFUN
$$
We have also derived the same result as eq.~\eqPGBETAFUN\ by an
independent calculation in the pure-gauge model of our non-linear
gauge without using eq.~\eqBETAFUN.

We assume that $\alpha$, $\beta$, $t^2$ and $u^2$ are all positive
quantities.
This is natural in order for the gauge fixing \eqOURLGAUGE\ to make
sense. $\beta_{1/t^2}$ and $\beta_{1/ u^2}$ in eq.~\eqPGBETAFUN\
of the pure-gauge model are consistent with this assumption since
we have
$\beta_{t^2}=0$ and $\beta_{u^2}=0$ at $t^2=0$ and $u^2=0$,
respectively.
In the Yang-Mills theory, $\beta_\alpha$ and $\beta_\beta$
in eq.~\eqBETAFUN\ tell that $\alpha$ and $\beta$ never flow
outside the region $\alpha >0$ and $\beta >0$
as $\mu$ is decreased starting from the inside of this region.

\section{Renormalization Group Flow in the Pure-Gauge Model}


\FIG\figPGETAZETAABEL{The RG flow of the variables $\eta$ and
$\zeta$ in the Abelian gauge subspace $\xi =1$ in the
pure-gauge model. The solid curves/lines and the broken ones are
those of $\beta_\zeta =0$ and $\beta_\eta =0$, respectively,
and the circle \circle indicates the fixed point.
The arrows are typical flow vectors
$(-\beta_\eta ,-\beta_\zeta )_{\xi =1}$
(up to norm) to the IR direction in each region bounded by the
$\beta_\zeta =0$ and $\beta_\eta =0$ curves/lines.
The shaded region is the region where the Abelian gauge is
IR unstable in the $\xi$-direction.}

\FIG\figPGXIZETA{The RG flow in the subspace $\eta =0$ in the
pure-gauge model.
The solid lines and the broken ones are those of $\beta_\zeta =0$
and $\beta_\xi =0$.
The arrows are typical flow vectors
$(-\beta_\xi ,-\beta_\zeta )_{\eta =1}$
(up to norm) to the IR direction in each region bounded by the above
lines.
The points indicated by \circle, $\triangle$ and $\times$ are the IR
sink, the saddle point and the IR source, respectively. }


Having obtained the $\beta$-functions our next task is to study the
RG flow to see whether the Abelian gauge is IR stable or unstable.
In this subsection we shall examine the (simpler) pure-gauge
model. The Yang-Mills theory will be studied in the next subsection.

First recall that the Abelian gauge $\xi=1$ is a fixed subspace of the
RG since we have $\beta_\xi =0$ at $\xi=1$, which is ensured by the
local $U(1)_3$ symmetry \eqUTHREEROT.
Note also that our gauge fixing space
is divided, by the signs of $\eta$ and $\zeta +1$, into four regions,
$$
\eqalign{
&I_{-+}:(\eta <0,\zeta \geq -1)\ ,
\qquad I_{++}:(\eta \geq 0,\zeta \geq -1)\ , \cr
&I_{--}:(\eta <0,\zeta <-1)\ ,
\qquad I_{+-}:(\eta \geq 0,\zeta <-1)\ , \cr
}
\eqn\eqFOURREGION
$$
which are not connected with each other by RG flows.
This is because we have $\beta_\eta =0$ and $\beta_\zeta =0$ at
$\eta =0$ and $\zeta =-1$,
respectively, owing to the symmetries of eq.~\eqRESTRICTION.
Since $\beta_\xi$ near the Abelian gauge is given by
$$
{1\over t^2}\beta_\xi \sim
{1\over 4}(\zeta +5)(1-\xi )\quad (\xi \sim 1) \ ,
\eqn\eqPGBETAXIABEL
$$
any point close to the Abelian gauge and lying in the region
$\zeta >-5$ ($\zeta <-5$) is repelled from (attracted to) the
Abelian gauge as it flows to the IR direction.
This implies that the Abelian gauge is IR unstable in the regions
$I_{++}$ and $I_{-+}$.

To see whether the Abelian gauge is IR stable or unstable in the other
two regions $I_{--}$ and $I_{+-}$, note first that $\beta_\eta$
and $\beta_\zeta$ on the Abelian gauge $\xi =1$ are given by
$$
{1\over t^2}\beta_\eta \Big\vert_{\xi =1}
=2\eta \left(\eta +{3\over 4}\right)(1+\zeta ) \ ,
\qquad
{1\over t^2}\beta_\zeta \Big\vert_{\xi =1}
=(1+\zeta )\bigl[\zeta +2\eta (1+\zeta )\bigr] \ .
\eqn\eqPGBETAABEL
$$
and hence the RG analysis in the Abelian gauge subspace is closed
within the two variables $\eta$ and $\zeta$ (the common factor
$t^2$ in eq.~\eqPGBETAABEL\ does not affect the shape of the
RG trajectory in the $(\eta ,\zeta )$ plane).
In order for the Abelian gauge to be
IR stable in the original parameter space
$(\xi ,\eta ,\zeta ,t^2,u^2)$,
there must be an IR stable fixed point in the region $\zeta <-5$
in the RG flow of the parameters $\eta$ and $\zeta$ restricted to the
Abelian gauge subspace. This RG flow
is given in fig.~\figPGETAZETAABEL\ and the arrows there indicate
the direction of the flow vector
$(-\beta_\eta ,-\beta_\zeta )\vert_{\xi =1}$
to the IR direction at various points in the $(\eta ,\zeta )$ space.
In fig.~\figPGETAZETAABEL\ there are two fixed points $(0,0)$
and P:$(-3/4,-3)$ satisfying $\beta_\eta =\beta_\zeta =0$
and they are both IR sinks in the $(\eta ,\zeta )$ plane.
(Any point on the line $\zeta =-1$ is also a fixed point
(it never moves) but is IR unstable in the $\zeta$-direction.)

{}From fig.~\figPGETAZETAABEL, we see that the Abelian gauge is IR
unstable in the region $I_{--}$. Any point on the Abelian gauge and
in the region $I_{--}$ is attracted to the fixed point P:$(-3/4,-3)$
lying in $\zeta >-5$.
For example, the RG trajectory starting from the point A in
fig.~\figPGETAZETAABEL\ is as drawn in the figure.
On the other hand, the Abelian gauge may have IR stable fixed points
in the region $I_{+-}$; at $(\eta =+\infty ,\zeta =-\infty )$
and $(\eta =0,\zeta =-\infty )$.

Let us look more carefully at the behavior of the RG flow near the
infinity. Making the approximation $\abs{\eta },\abs{\zeta }\gg 1$
in the $\beta$-functions \eqPGBETAXIABEL\ and \eqPGBETAABEL,
the solution of the RG equation in the regions close to the Abelian
gauge and also close to
$(\eta ,\zeta )=(+\infty ,-\infty )$ or
$(-\infty ,+\infty )$ is given by
$$
\zeta =-a\eta \ ,\quad t^2=b/\abs{\eta }\ ,\quad
\abs{1-\xi }=c\exp\!\left({1\over 8\eta }\right)\ ,\quad
\abs{\eta }=16\pi^2ab/\ln\left(\mu_0/\mu \right)\ ,
\eqn\eqASYMPSOL
$$
where $a$, $b$, $c$ $(>0)$ and $\mu_0$ are integration constants.
Note that $\abs{1-\xi }$ does not vanish (diverge) even at infinity
$\eta \rightarrow +\infty $ ($-\infty $),
and hence the Abelian gauge is not sufficiently IR
stable (unstable) there.
However, in any case, the one-loop $\beta$-function is totally
unreliable in these regions: for example, the strength of the
$\cb^3 c^3\cb^\pm c^\mp$ interaction in eq.~\eqOURLGAUGE,
$t^2(1+\zeta )\eta$, diverges as
$t^2\zeta \eta \sim \eta \rightarrow \infty$.

It would be very interesting if we could compactify the
$(\eta ,\zeta )$ space
(into torus?) to continue the RG trajectory extending to the infinity
$(\eta =+\infty ,\zeta =-\infty )$ to another infinity,
$(\eta =-\infty ,\zeta =-\infty )$ or
$(\eta =+\infty ,\zeta =+\infty )$.
If this is possible, the Abelian gauge can be totally IR unstable
in all four regions $I_{\pm,\pm}$.
However, this is beyond the scope of perturbation theory.

Here we summarize the IR stability/instability of the Abelian gauge in
the pure-gauge model. The Abelian gauge is IR {\it unstable} in the
three regions $I_{++}$, $I_{-+}$ and $I_{--}$.
The Lorentz gauge with $(\eta ,\zeta )=(0,0)$ or $(-1,-1)$ and the
$OSp(4/2)$ gauge with $(-1/2,-1/2)$ is contained in these regions.
In the other region $I_{+-}$, the Abelian gauge may have IR stable
fixed points at the infinity of the $(\eta ,\zeta )$ space.
There is also a (non-perturbative) possibility that the Abelian gauge
is globally IR unstable if we are allowed to regard the
$(\eta ,\zeta )$ space as compactified.

Next, let us discuss the RG trajectory more globally away from the
Abelian gauge. The complete analysis will require numerical treatment.
Here we shall content ourselves with partial analysis.
First let us consider the fixed points having the global $SU(2)$
symmetry. Putting the $SU(2)$ symmetry condition $\xi =0$,
$\eta =\zeta$, and $r\equiv t^2/u^2=1$ in the
$\beta$-functions we have
$$
\eqalignno{
&{1\over t^2}\beta_\eta \Big\vert_{SU(2)}
={1\over t^2}\beta_\zeta \Big\vert_{SU(2)}
= 2\zeta \left(\zeta  + \Half \right)(\zeta +1) \ ,
&\eqname{\eqPGBETAZETASU} \cr
&\beta_{1/t^2}\Big\vert_{SU(2)}=\beta_{1/u^2}\Big\vert_{SU(2)}
= 1 + 2\zeta (1+\zeta ) \ ,
&\eqname{\eqPGBETATSU} \cr
}
$$
and $\beta_\xi =\beta_r=0$ is automatically satisfied.
Eq.~\eqPGBETAZETASU\ tells that in the $SU(2)$ symmetric subspace
the Lorentz gauge points $\zeta =0$ and $-1$ are IR stable while the
$OSp(4/2)$-symmetric gauge point $\zeta =-1/2$ is IR unstable.
Analysis in the original coupling constant space
$(\xi ,\eta ,\zeta ,r)$
shows that the Lorentz gauge points, $(0,0,0,1)$ and $(0,-1,-1,1)$,
are IR stable in all directions while the $OSp(4/2)$-symmetric gauge
$(0,-1/2,-1/2,1)$ is IR stable in three directions and unstable in one
direction. (However, this result need not be taken so seriously since
the coupling constants $t^2$ and $u^2$ diverge at these fixed points
as seen from eq.~\eqPGBETATSU.)
As mentioned in the Introduction, the pure-gauge model with
$OSp(4/2)$-symmetric gauge is (in the Parisi-Sourlas sense) exactly
equivalent to the (solvable) chiral model in two dimensions, and hence
the KO condition \eqKOCOND\ is known to be satisfied. Unfortunately
this $OSp(4/2)$-symmetric gauge is not IR stable in all directions.
However, this is not a bad news since we expect that the Lorentz gauge
pure-gauge model, which also shares common properties with the chiral
model in two dimensions, also satisfies the KO condition \eqKOCOND.

Finally, as an example of the RG flow away from the Abelian gauge, let
us consider the $\eta =0$ subspace which is a fixed subspace of the RG.
On this subspace we have
$$
{1\over t^2}\beta_\xi \Big\vert_{\eta =0}
= {1\over 4}(1-\xi )(5\xi +\zeta )\ ,\quad
{1\over t^2}\beta_\zeta \Big\vert_{\eta =0} = (1+\zeta )\zeta  \ ,
\eqn\eqPGBETAXIZETA
$$
and the analysis is closed within the two variables
$\xi$ and $\zeta$.
Fig.~\figPGXIZETA\ shows the flow vector
$(-\beta_\xi ,-\beta_\zeta )$
(up to norm) to the IR direction at various points in the
$(\xi ,\zeta )$ space. In fig.~\figPGXIZETA\ there are four
fixed points satisfying $\beta_\xi =\beta_\zeta =0$ :
$(\xi ,\zeta )= (0,0)$ is an IR sink, $(1,0)$ and $(1/5,-1)$ are
saddle points, and $(1,-1)$ is an IR source.
The origin $(0,0)$ represents the Lorentz gauge if another condition
$r=1$ is satisfied. This is the case in the IR since we have
$$
{1\over t^2}\beta_r\Big\vert_{\eta =0}\sim \Half (r-1)\quad
\left((\xi ,\zeta )\sim (0,0)\right) \ ,
\eqn\eqPGBETARZEROZERO
$$
which implies that $r$ is attracted to $r=1$ as $\mu$ is decreased.
The presence of the fixed points $(0,0)$ and $(1,-1)$ are ensured by
the symmetries (\cf\ \eqRESTRICTION) but the other two fixed points
might be the artifacts of the one-loop calculation.

Fig.~\figPGXIZETA\ tells the followings. First, in the upper region
$\zeta >-1$, the Abelian gauge is IR unstable,
and any point $(\xi ,\zeta )$
flows either to the Lorentz gauge $(0,0)$
(when $\xi <1$) or to the infinity $(\xi =+\infty ,\zeta =0)$
(when $\xi >1$).
We do not know a definite answer to ``save" the flow to the infinity.
One possibility would be to compactify the $\xi$ space. Another
possibility would be to relax the charge conjugation symmetry
eq.~\eqPARITY\ to consider a larger gauge fixing space eq.~\eqGENERALG.
Then a point in the region $(\xi >1,\zeta >-1)$ having small
values of the new coupling constants might be attracted to the Lorentz
gauge in the IR. In fact, a preliminary calculation by complexifying
the parameter $\xi$ supports this expectation.

As for the lower region $\zeta <-1$ of fig.~\figPGXIZETA,
the RG trajectory starting from the neighborhood of the Abelian gauge
is consistent with the flow on the $\eta =0$ line of
fig.~\figPGETAZETAABEL: though it first deviates from $\xi =1$ in the
region $-5<\zeta <-1$, the trajectory extends to the infinity
$(\xi =1,\zeta =-\infty )$ on the Abelian gauge.
Note, however, that the subspace $\eta =0$ we are considering is IR
unstable in the region $\zeta <-1$ near the Abelian gauge, and the
above flow is valid only just on the $\eta =0$ subspace.


\section{Renormalization Group Flow in the Yang-Mills theory}

\FIG\figETAZETAABEL{The RG flow in the Abelian gauge subspace $\xi =1$
projected on the $(\eta ,\zeta)$ plane for various values of $\alpha$
in the Yang-Mills theory.
The range of $\alpha$ of the four figures are
(A): $\alpha <\left(\sqrt{51}-6\right)/5$,
(B): $\left(\sqrt{51}-6\right)/5<\alpha <\sqrt{2}-1$,
(C): $\sqrt{2}-1<\alpha <\sqrt{3}$ and
(D): $\alpha >\sqrt{3}$.
The solid curves/lines and the broken ones are those of
$\beta_\zeta =0$ and $\beta_\eta =0$, respectively,
and the circle \circle indicates the
``fixed point" with $\beta_\zeta =\beta_\eta =0$.
The arrows are typical flow vectors
$(-\beta_\eta ,-\beta_\zeta )_{\xi =1}$
(up to norm) in each region bounded by the above curves/lines.
The Abelian gauge is IR unstable in the $\xi$-direction in the shaded
region. The vertical scales are different among the four figures.
}


The RG analysis for the Yang-Mills theory is more involved than in the
pure-gauge model but is similar.
First, we see immediately that the Abelian gauge is IR unstable in the
regions $I_{++}$ and $I_{-+}$ of eq.~\eqFOURREGION. This is because
$\beta_\xi$ near the Abelian gauge is given by
$$
{1\over g^2}\beta_\xi \sim {1\over 4}\left(
 12 + \alpha (5+\zeta ) \right)(1-\xi ) \quad (\xi \sim 1) \ ,
\eqn\eqBETAXIABEL
$$
and we have $12 +\alpha (5+\zeta)>0$ in these regions
(recall that $\alpha >0$).
In order to examine the IR stability of the Abelian gauge in the other
two regions $I_{--}$ and $I_{+-}$, let us consider the RG flow on the
Abelian gauge subspace. In the case of Yang-Mills theory we have to
consider the three dimensional coupling constant space
$(\eta ,\zeta ,\alpha )$
since the $\beta$-functions on the Abelian gauge are given by
$$
\eqalign{
&{1\over g^2}\beta_\zeta \Big\vert_{\xi =1} =
\alpha (1+\zeta )\left\{\zeta +{3\over \alpha^2} +
2\eta (1+\zeta )\right\} \ ,\cr
&{1\over g^2}\beta_\eta \Big\vert_{\xi =1} =
\eta \left\{ 3\left(1+{1\over \alpha }\right) +
\alpha \left({3\over 2}+2\eta \right)(1+\zeta ) \right\} \ ,\cr
&{1\over g^2}\beta_\alpha \Big\vert_{\xi =1} = \alpha \left\{
{4\over 3} - \alpha \left(
2 + \zeta  + 2\eta (1+\zeta )\right) \right\} -3 \ .\cr
}
\eqn\eqBETAABEL
$$
Fig.~\figETAZETAABEL\ shows the RG flow vector
$(-\beta_\eta ,-\beta_\zeta )_{\xi =1}$ in the $(\eta ,\zeta)$ space
at various values of $\alpha$.
In the present case we have to take into account
that the value of $\alpha$ is also varied as the renormalization
point $\mu$ is decreased in reading the RG flow from
fig.~\figETAZETAABEL.

{}From fig.~\figETAZETAABEL, we see first that the Abelian gauge is
IR unstable in the region $I_{--}$.
One might be afraid that (differently from the pure-gauge model case)
there is a ``fixed point" P:$(0,-3/\alpha^2)$ satisfying
$\beta_\eta =\beta_\zeta =0$ in the Abelian gauge stable region
$12 +\alpha (5+\zeta )<0$
when $\alpha <\left(\sqrt{51}-6\right)/5$ (fig.~\figETAZETAABEL~(A)).
However, this ``fixed point" is an IR source in the
$(\eta ,\zeta )$ plane
and any point near there having a negative $\eta$ coordinate flows to
the Abelian gauge unstable region $12 +\alpha (5+\zeta )>0$.

Now we are left with the region $I_{+-}$.
First note that the situation here is ``better" than in the pure-gauge
model case since when $\alpha <\sqrt{3}$, there is a region
(the region with dots in figs.~\figETAZETAABEL\ (A), (B) and (C))
where the flow vector points upward.
Starting from a point in the region $I_{+-}$ with large $\eta$ and
$\abs{\zeta }$, let us trace its RG trajectory to the IR direction.
In this region $\alpha$ decreases as $\mu$ is decreased, and
the point P:$(0,-3/\alpha^2)$ moves downward.
Therefore, there might be a possibility that the RG trajectory is
outstripped by the curve surrounding the dotted region,
and the trajectory begins moving upward and finally escapes
from the shaded region.
If this is true, the Abelian gauge is IR unstable also in the
region $I_{+-}$. However, we cannot confirm this possibility since
perturbation theory is unreliable in the region with large $\eta$ and
$\abs{\zeta }$.

As for the IR stability/instability of the Lorentz gauge and the
$OSp(4/2)$ gauge, it is the same as in the pure-gauge case.
In the present case of Yang-Mills theory, we have to consider the five
dimensional space $(\xi,\eta,\zeta,r,\alpha)$ since they are all
coupled in the RG equations.
For both the Lorentz gauge and the $OSp(4/2)$ gauge there are two
possibilities for the value of $\alpha$: $\alpha =+\infty$ or $0$.
In both cases
the Lorentz gauge is an IR sink in all directions while the $OSp(4/2)$
gauge is IR stable in four directions and unstable in one direction.

\chapter{Summary and discussion}

We have found that (in the one-loop approximation) the Abelian gauge
is IR unstable at least in three of the four regions of the gauge
parameter space that are not connected by RG flows.
The Lorentz gauge is contained in these three regions and it is an IR
sink.
These results are common to the (pure) Yang-Mills theory and the
pure-gauge model.
We have pointed out the possibility that the Abelian gauge is also
IR unstable in the fourth region.
Although our analysis is based on the one-loop $\beta$-functions,
we expect that the conclusion of this paper, that the Abelian gauge
is IR unstable at least in the three regions, remains valid even if
exact treatments become possible.
Actually in drawing our conclusion we do not need to consider the far
IR region where the coupling constants blows up .

That the Abelian gauge is IR unstable means that the infrared
properties of the theories with $\xi \equiv 1$ and $\xi \not=1$ are
totally different and therefore the breakdown of the KO confinement
mechanism may be a phenomenon which occurs only in the Abelian gauge.
Concretely, we would have
$\lim_{p^2\rightarrow 0}\lim_{\xi \rightarrow 1}u^{33}(p^2)=0$
but
$\lim_{\xi \rightarrow 1}\lim_{p^2\rightarrow 0}u^{33}(p^2)=-1$.
Since $u^{33}(p^2)$ is proportional to $1-\xi$, a possible
$u^{33}(p^2)$ for small $p^2$ would be
$u^{33}(p^2)=-(1-\xi )/\left(ap^2 + 1-\xi \right)$.

If we conclude that the Abelian gauge is IR unstable, we are left with
the problem: why is the KO mechanism inapplicable for
explaining color confinement in the Abelian gauge?
There must be something special about the Abelian gauge that renders
the application of the KO mechanism meaningless.
One possibility would be that the gauge field
(in particular, $A^3_\mu$) is an ``angle variable" in the Abelian
gauge, by which we mean the following:
The Abelian gauge is characterized as having the local gauge symmetry
\eqUTHREEROT\ with parameter $\epsilon^3(x)=a_\mu x_\mu$.
However, this symmetry is {\it always} spontaneously broken and
$A^3_\mu$ is the corresponding Nambu-Goldstone boson.
On the other hand, in the Lorentz gauge, the KO condition is
interpretable as the restoration of a similar kind
of local gauge symmetry as was stated in the
Introduction\rlap.\refmark{\HataRES}
The restoration is possible since the parameter is field dependent.
If we make an analogy between the choice of gauge in the Yang-Mills
theory and the choice of the parametrization of the group element in
the chiral model,
we know a similar phenomenon in the chiral model.
Consider, for example, the $SU(2)\times SU(2)$ chiral model in two
dimensions and take the following parametrization of the $SU(2)$ group
element $g(x)$:
$g(x)=\exp\left\{\sumpm(\sigma^\mp/2i)\varphi^\pm (x)\right\}
\exp\left\{(\sigma^3/2i)\varphi^3(x) \right\}$.
Then, the right transformation around the 3rd axis,
$g\rightarrow g\exp\left((\sigma^3/2i)\epsilon^3\right)$, with a
constant angle $\epsilon^3$
is expressed in terms of $(\varphi^\pm,\varphi^3)$ as
$(\varphi^\pm,\varphi^3)\rightarrow
(\varphi^\pm,\varphi^3+\epsilon^3)$,
and this symmetry seems always to be spontaneously broken, although
the $SU(2)\times SU(2)$ symmetry should be fully linearly
realized in this model.
This apparent contradiction is because $\varphi^3$ is an angle variable
with period $4\pi$ and it is meaningless to use an angle variable
itself in deciding the spontaneous symmetry breaking.
What we expect in the case of Yang-Mills theory in the Abelian gauge
is that the field $A^3_\mu$ is an angle variable in some sense.
This possibility will be explored elsewhere\rlap.\refmark{\HataIkkiII}

Finally we wish to point out the possibility of applying the KO
mechanism to string theory, and in particular, to string field
theory\refmark{\SFT} for explaining the expected dynamical reduction
of the degrees of freedom at the Planck scale\rlap.\refmark{\Planck}
In fact, string field theory is a kind of gauge theory having an
infinite number of local fields and associated local gauge
symmetries, and is quantized using the BRST method.
Therefore, if the KO mechanism works for all the gauge
symmetries, most of the degrees of the freedom in string field
theory would become unphysical.
Recalling the interpretation of the KO mechanism as the restoration
of a local gauge symmetry\rlap,\refmark{\HataRES}
an infinite number of local gauge symmetries in string field theory,
most of which are spontaneously broken in perturbation theory,
should be restored at short distances.
To pursue this possibility we have to first resolve the
conceptual problem of reconciling the ``short-distance scale"
with the ``asymptotic fields" in the KO mechanism.

\ACK

We would like to thank the members of the particle theory group at
Department Physics, Kyoto University, especially, T.~Kugo,
D.~Lancaster, T.~Maskawa and M.G.~Mitchard for valuable discussions
and useful comments.
We also wish to thank M.G.~Mitchard for careful reading of the
manuscript.

\APPENDIX{A.}{A}

In this Appendix we reproduce the proof that the BRST exact expression
of the color charge, eq.~\eqQA, implies color confinement in the sense
stated below eq.~\eqQA.
Here we prove the equivalent statement: under the assumption that
eq.~\eqQA\ is true, any physical asymptotic state is
necessarily a color singlet.

Let $a^\dagger_i$ be the creation operator of a physical asymptotic
field belonging to some representation of color $Q^a$
($i$ is the color index).
{}From the assumption that it is physical and belongs to a representation
of color, we have
$$
[\QB, a^\dagger_i]=0\ , \quad
[a_i, a^\dagger_j]=\delta_{ij}\ , \quad
[Q^a, a^\dagger_i ]=a^\dagger_j T^a_{ji}\ ,
\eqn\eqPHYSICAL
$$
where $T^a_{ij}$ is the color matrix in some representation.
Then, from $\bra{0}a_iQ^a a^\dagger_j\ket{0}=0$, which is a consequence
of the BRST exact expression of $Q^a$, eq.~\eqQA, and $\QB\ket{0}=0$,
we can show that $T^a_{ij}=0$:
$$
0 = \bra{0}a_i Q^a a^\dagger_j\ket{0} =
\bra{0}a_i [Q^a,a^\dagger_j ]\ket{0} =
\bra{0}a_i a^\dagger_k\ket{0} T^a_{kj} = T^a_{ij} \ ,
\eqn\eqZERO
$$
where use has been made of $Q^a\ket{0}=0$.
Therefore, any physical asymptotic field must be a color singlet.

\refout

\figout

\bye